      [1999/12/01 v1.4c Il Nuovo Cimento]
\documentclass{cimento}
\usepackage{epsfig}
\usepackage{graphics}

\title{GRB 060117: Reverse + forward shock solution}

\author{Martin~Jel\'{\i}nek\from{ins:iaa}
\thanks{mates@iaa.es},
Michael Prouza\from{ins:col}\from{ins:fzu}\from{ins:aug}, Petr~Kub\'anek\from{ins:asu},\\
Ren\'{e}~Hudec\from{ins:asu},
Martin~Nekola\from{ins:asu}, Jan \v{R}\'{\i}dk\'{y} \from{ins:fzu}\from{ins:aug}
\atque Ji\v{r}\'{\i}~Grygar\from{ins:fzu}\from{ins:aug}
}
\shortauthor{M. Jel\'{\i}nek, M. Prouza, P. Kub\'anek et al.}
\instlist{\inst{ins:iaa} Instituto de Astrof\'\i sica de Andaluc\'\i a,
180\,08 Granada, Spain
  \inst{ins:col} Nevis Institute and Department of Physics, Columbia University, New York, NY 10027, USA
  \inst{ins:fzu} Fyzik\'{a}ln\'{\i} \'ustav AV \v{C}R, Na Slovance 2, Praha 8, 
  \inst{ins:asu} Astronomick\'{y} \'ustav AV \v{C}R, Fri\v{c}ova 298,
  Ond\v{r}ejov, Czech Republic
  \inst{ins:aug} for the Pierre Auger Collaboration
}
\PACSes{
\PACSit{95.55.Cs}{Ground based ultraviolet, optical and infrared telescopes}
\PACSit{95.75.De}{Photography and photometry}
\PACSit{98.70.Rs}{Gamma-ray sources; gamma-ray bursts}
}

\newcommand{\swift} {{\it Swift}}

\begin{document}

\maketitle

\begin{abstract} 
We present a discovery and observation of an extraordinarily
bright prompt optical emission of the GRB\,060117 obtained by a
wide-field camera atop the robotic telescope FRAM of the Pierre Auger
Observatory from 2 to 10 minutes after the GRB.
We found rapid average temporal flux decay of $\alpha = -1.7 \pm 0.1$
and a peak brightness $R = 10.1$ mag.
We interpret the shape of the lightcurve 
as a transition between reverse and forward shock emission.
\end{abstract}

\section{Swift detection}

A bright long-soft GRB\,060117 was detected by \swift\/ satellite on
January 17, 2006, at 6:50:01.6\,UT.
It showed a multi-peak structure with T$_\mathrm{90}$=$16\pm1$\,s with
maximum peak flux $48.9\pm1.6\,{\rm ph\,cm}^{\rm -2} {\rm s}^{\rm -1}$. Thus, 
GRB060117 was --- in terms of peak flux --- the most intense GRB detected so far by Swift
Coordinates computed by \swift\/ were available within 19\,s and
immediately distributed by GCN \cite{gcn4538}.

\section{FRAM and optical transient observation}

FRAM is part of the Pierre Auger cosmic-ray observatory \cite{auger},
and its main purpose is to immediately monitor the atmospheric
transmission.  
FRAM works as an independent, RTS2-driven, fully robotic
system, and it performs a photometric calibration of the sky on various
UV-to-optical wavelengths using a 0.2\,m telescope and a photoelectric
photomultiplier. 
%

FRAM received the notice at 06:50:20.8\,UT, 19.2\,s after the trigger
and immediately started the slew. 
The first exposure started at 06:52:05.4, 123.8\,s after the GRB. 
Eight images with different exposures were taken before the observation
was terminated. 
A bright, rapidly decaying object was found, and its presence was
reported by \cite{gcn4535} soon after the discovery.
The FRAM lightcurve for this optical transient is in Figure 1. 

\begin{figure}[t!] 
        \begin{center} \label{fig4}
        \resizebox{0.4\hsize}{!}{
	\includegraphics{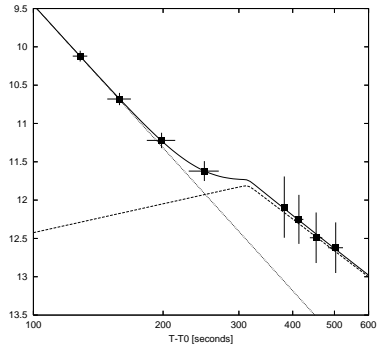}
	} 
        \caption{\small The R-band afterglow lightcurve of GRB\,060117. 
        The lightcurve is fitted as a superposition of reverse shock
        (dotted line) and forward shock (dashed line).
%
	}
        \end{center} 
\end{figure}

\section{Interpretation}

Our preffered interpretation (based on the work of \cite{shao05}) is to fit the 
data as a 
transition between the reverse and the forward shock with the passage of
the typical frequency break $\nu_m$ through the observed passband at
time $t_{m,f}$. 
%
Corresponding decay indices are $\alpha_{\mathrm Reverse}$=2.49$\pm$0.05 and
$\alpha_{\mathrm Forward}$=1.47$\pm$0.03 (see Fig\,3).

Other possible interpretations and more details about FRAM telescope, data processing and other
follow-up attempts can be found in \cite{aal}.

\acknowledgments
{\small The telescope FRAM was built and is operated under the support of the
Czech Ministry of Education, Youth, and Sports through its grant
programs LA134 and LC527.
MJ would like to thank to the Spanish Ministry of Education and Science
for the support via grants AP2003-1407, ESP2002-04124-C03-01, and
AYA2004-01515 (+ FEDER funds), MP was supported by the Grant Agency of
the Academy of Sciences of the Czech Republic grant B300100502.}

\end{document}